\documentclass[preprint,showpacs,preprintnumbers,amsmath,amssymb]{revtex4}

\usepackage{graphicx}
\usepackage{dcolumn}
\usepackage{bm}

\begin{document}

\title{The many origins of charge inversion in electrolyte solutions:
effects of discrete interfacial charges}

\author{J. Faraudo,$^{1}$ and A. Travesset$^{2}$}
\affiliation{$^{1}$Departament de Fisica, Universitat Autonoma de Barcelona, Bellaterra, Spain;\\
$^{2}$ Department of Physics and Astronomy and Ames Laboratory,
Iowa State University, Ames, Iowa 50011}

\date{\today}

\begin{abstract}

We show that charge inversion, i.e. interfacial charges attracting
counterions in excess of their own nominal charge, is a general
effect that takes place in most charged systems next to aqueous
solutions with multivalent ions and identify three different
electrostatic origins for this effect 1) counterion-counterion
correlations, 2) correlations between counterions and interfacial
charges and 3) complexation. We briefly describe the first two
regimes and provide a detailed characterization of the
complexation regime from united atom molecular dynamics simulation
of a phospholipid domain in contact with an aqueous solution. We
examine the expected conditions where each regime should apply and
describe a representative experimental example to illustrate each
case. We point out that our results provide a characterization of
ionic distributions irrespectively of whether charge inversion
takes place and show that processes such as proton release and
transfer are also linked to ionic correlations. We conclude with a
discussion of further experimental and theoretical implications.
\end{abstract}

 \maketitle

\section{Introduction}\label{SECT__Intro}

The electrostatics of molecules in aqueous media is key for
understanding fundamental biological or physico-chemical processes
and exhibits a fascinatingly diverse range of phenomena that
remains the subject of intense theoretical and experimental work
\cite{Boroud05,Grosberg2002,Levin2002}. A relevant example is
charge inversion, where interfacial charges attract counterions in
excess of their own nominal charge, thus leading to an interface
whose effective charge is opposite in sign. Experimental examples
of charge inversion have been observed on a wide range of systems
such as lipid vesicles, solid interfaces, colloids or Langmuir
monolayers among others, in contact with an aqueous solution
containing multivalent ions
\cite{McLaughlin1989,Besteman2004,Besteman2005,Israelachvili2000,Vaknin2003,Pitler2006}.
At the theoretical level, although there is consensus that the
origin of charge inversion lies in the presence of correlations
among charged objects in solution \cite{Grosberg2002}, it is still
unclear to what extent the different proposed theories describe
the experimental data.

With few exceptions (see
\cite{Moreira2002,Lukatsky2002,Henle2004,Travesset2005,Travesset2006}),
previous theoretical studies have assumed that interfacial charges
can be smeared to a uniform distribution. In this paper, we
investigate effects related to the discrete nature of interfacial
charges and their possible conformational degrees of freedom and
further elucidate how they lead to charge inversion. Although, as
stated, the subject of this paper is mainly charge inversion, this
effect is a consequence of the spatial distribution of charges in
solution, so an elucidation of the former provides a detailed
understanding for the latter. Therefore, the results presented in
this paper provide a precise characterization of ionic
distributions next to charged interfaces, which applies also in
those cases where charge inversion does not take place.

The organization of the paper is as follows. In
Sect.~\ref{SECT__Theory} we discuss the different correlation
regimes and describe their main features. A comparison of new
simulation results with the different theories is provided in
Sect.~\ref{SECT__Sim}, which also provides a detailed example for
the newly introduced complexation regime. The discussion of the
expected range of validity of the different correlation regimes
and its relation to existing experiments is provided in
Sect.~\ref{SECT__Discussion}. We finish with some general
conclusions in Sect.~\ref{SECT__Conclusions}.

\section{Correlation regimes and charge inversion.}
\label{SECT__Theory}

In the analysis of the electrostatics of charged interfaces in
contact with electrolyte solutions it is customary to divide the
aqueous solution into two regions, the so-called Stern and diffuse
layers \cite{Israelachvili2000}. The Stern layer corresponds to
the region immediate to the interface and may include bound
counterions whereas the diffuse layer consists of an atmosphere of
ions in rapid thermal motion. A common phenomenological approach
describes the Stern layer as a Langmuir isotherm with several
phenomenological parameters (such as binding constants,
interfacial dielectric constant for water, etc..) coupled with the
classical Poisson-Boltzmann (PB) \cite{Safran1994} theory
describing the diffuse layer. While the description of the diffuse
layer by PB theory is usually quite satisfactory (almost exact in
the dilute regime), a more rigorous description of the Stern layer
without resorting to empirical parameters presents considerable
theoretical difficulties, but this description is necessary for an
unambiguous description of the correlations that lead to charge
inversion.

Let us illustrate in some detail how correlations induce charge
inversion \cite{Grosberg2002}. We will assume solutions with
approximately unit activity, so that the chemical potential of the
counterions within the diffuse layer is simply $\mu_{diff}=k_B T
\ln(n_Cv_0)$, where $n_C$ is the salt number density of the
solution and $v_0$ can be taken as the volume of a single
counterion (we note that more concentrated regimes can be
described from the results in \cite{Pianegonda2005}). The chemical
potential within the stern layer is given by $\mu_{Stern}=k_B T
\ln(n_C^s s_0)+q_C e\phi_0+\mu_{corr}$, where $n_C^s$ is the two
dimensional number density of counterions with cross-sectional
area $s_0$ and ionic valence $q_C$, while $\phi_0$ is the contact
value potential and $\mu_{corr}$ is the contribution arising from
correlation effects. The onset of charge inversion takes place at
a salt concentration $c_{inv}$, where the contact value potential
is zero ($\phi(0)=0$) and the Stern layer number density satisfies
the neutrality condition $n_C^s=-\sigma_0/q_Ce$, where $\sigma_0$
is the ``bare''interfacial charge density. This leads to the
equation
\begin{equation} \label{eq:inv_general}
     c_{inv}=\frac{|\sigma_0|}{2r_C q_C e}\exp\left( \frac{\mu_{corr}}{k_BT} \right),
\end{equation}
where $r_C$ is the counterion radius. For charge inversion to
occur within the dilute regime (ionic strengths of the order of
0.1M or lower) a significant favorable correlation energy is
necessary (more precisely, the correlation chemical potential must
satisfy $\mu_{corr}<< -k_BT $).

Where the favorable correlation energy comes from? if electric
charges are smeared into a continuum, we revert back to the
standard PB description where $\mu_{corr}=0$ and no charge
inversion can take place. In other words, a favorable correlation
energy arises from effects that are related to the underlying
discreteness of electric charges. We next discuss different
explicit scenarios that lead to $\mu_{corr}<< -k_BT$ and give
raise to charge inversion.

\subsection{Counterion-counterion or lateral Correlation (LC) regime}

This regime is dominated by counterion-counterion correlations
within the Stern layer, consequently approximating interfacial
charges as a smeared uniform charge density. This regime will be
designated as lateral correlation (LC) regime herein, and has been
studied quite extensively, so we refer to
Refs.~\cite{Shklovskii:99,Grosberg2002,Boroud05,Pianegonda2005}
for detailed presentations. Here we just review the most salient
features.

The magnitude of the counterion-counterion correlations is
quantified by the coupling parameter (or plasma parameter)
$\Gamma$ defined as:
\begin{equation}
 \Gamma = \frac{q^2_C e^2}{\varepsilon a_C  k_BT}=\frac{q^2_C l_B}{a_C},
\end{equation}
where $q_C$ is the valence of the counterions, $a_C$ is the
typical lateral counterion separation within the Stern layer and
$l_B=7.1$\AA \ is the Bjerrum length \cite{Safran1994}. At strong
coupling $\Gamma>>1$, the counterions form a strongly correlated
two dimensional liquid (the one component plasma (OCP)) that
provides the favorable free energy \cite{Totsuji1978} required for
charge inversion. The most relevant predictions for LC theories
are:
\begin{itemize}
\item The correlation chemical potential $\mu_{corr}$ is given by
the chemical potential of the OCP, quoted for example in
\cite{Totsuji1978}.
    \item The counterion pair distribution function within the Stern layer
    is described by the OCP (quoted in \cite{Totsuji1978}).
    \item Counterion distributions from the interface fall of
    exponentially \cite{Shklovskii:99,Moreira2001}
    \begin{equation}
    \label{eq:g_r_LC_O}
g_{\rm C I}\approx \exp(-\frac{z}{\lambda_{G}}) \ ,
\end{equation}
where $\lambda_{GC}$
    is the Guoy-Chapman length, which characterizes the counterion separation
    from the interface \cite{Safran1994}.
    \item Charge inversion occurs for large values of the parameter $\zeta$
defined as \cite{Grosberg2000}:
\begin{equation}
\label{eq:zeta} \zeta=q_{+}e/\pi \sigma_0 \lambda_D^2 \ ,
\end{equation}
where $\lambda_D$ is the Debye length.
\end{itemize}

\subsection{Counterion-interfacial charge or transverse correlation (TC) regime}

This regime is dominated by correlations between interfacial
charges and single counterions bound to the interfacial groups
with a fixed stoichiometric ratio. Herein, it will be designated
as the transverse correlation (TC) regime.

The description of the TC regime consists of an interface
providing a number of binding sites (with a binding constant
$K_I$) for counterions. In its simplest version, the Stern layer
is described as a Langmuir adsorption theory
\cite{Ninham1971,Healy1978,Travesset2006}. As an illustrative
example, we consider divalent cations $C^{++}$ with bulk number
density $n^B_C$ binding to an interface containing a surface
concentration {$[P^-]$} of singly charged molecules. Assuming that
each interfacial molecule provides a binding site for each
counterion (that is a 1:1 stoichiometric ratio) the interface
consists of a mixture of $[P^-]$ and $[C^{++}P^-]$ species in
equilibrium
\begin{equation}
\label{Langmuir} [ C^{++}P^-] / [ P^-] =K_I n_C(0)=K_I n^B_C
e^{-2\phi_0 /k_BT},
\end{equation}
where $n_C(0)$ and $\phi_0$ are respectively the contact value
counterion concentration and the contact value potential. The
surface charge density $\sigma$ at the Stern layer, which includes
the bound counterions, is given by
 \begin{equation}
 \label{sigma}
 \sigma / \sigma_0 = \frac{1-K_I n_C(0)}{1+K_I n_C(0)}=
\frac{1-K_I n^B_C e^{-2\phi_0/k_BT}}{1+K_I n^B_C e^{-2\phi_0/k_B
T}}
 \end{equation}
where $\sigma_0$ has already been defined as the bare surface
charge of the interface. This equation serves as a boundary
condition to the Poisson-Boltzmann equation \cite{Safran1994},
thus providing a self-consistent solution with $K_I$ as a free
parameter.

\begin{figure}
\includegraphics[width=8cm]{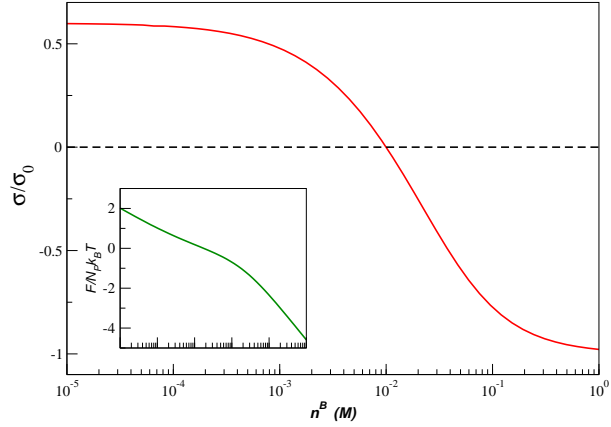}
\caption{Plot of the ratio of the effective charge of the
interface $\sigma$ to its bare charge $\sigma_0$ as a function of
bulk concentration. The results are for $\sigma_0=-e/80$
\AA$^{-2}$ and divalent counterions, with $K_L=100$ M$^{-1}$. The
onset of charge inversion $\sigma=0$ is at $c_{inv}=1/K_L$.
Complete reversal ($\sigma=-\sigma_0$) takes place at large
concentrations provided that counterion-counterion repulsion can
be ignored. The inset shows the free energy per particle as a
function of bulk concentration. } \label{fig:Inversion_TC}
\end{figure}

The condition for charge inversion is $\sigma/\sigma_0<0$, which
implies (see Eq.(\ref{sigma})) $n^B>1/K_I$. As an specific
example, we consider an interface consisting of singly charged
groups with a molecular area of $80$ \AA$^2$ (typical of a charged
phospholipid such as Phosphatidylserine) in contact with a
solution containing divalent counterions. The binding constant is
taken as $K_I=100$ M$^{-1}$ (A discussion of the experimental
relevant values is provided is Sect.~\ref{SECT__Discussion}). The
interfacial charge $\sigma$ is shown in
Fig.~\ref{fig:Inversion_TC} as a function of salt concentration.
The interface is neutralized at counterion concentrations
$c_{inv}=1/K_I=10^{-2}$M, and the charge is reversed for
concentrations larger than $c_{inv}$ (becoming completely
reversed, $\sigma=-\sigma_0$ for large enough counterion
concentrations).

If only electrostatic interactions are involved the binding
constant $K_I$ can be computed by generalizing the Bjerrum pairing
theory of electrolytes \cite{Robinson:1959}
\begin{equation} \label{eq:Bjerrum_constant}
    K_I^{TC}= c_{g} 4 \pi (|q_{+} q_{-}| l_B)^3 G(\frac{|q_{+}
    q_{-}|l_B}{r_0}),
\end{equation}
where  $G(x)=\int^x_2 dz z^{-4} e^z$, $r_0$ is the sum of the
crystallographic radius of the bound ions and the interfacial
charge and $q_{+}$,$q_{-}$ their valences. The factor $c_{g}<1$
takes into account steric constrains, with $c_{g}=1/2$ being a
reasonable value \cite{Travesset2006}. The binding constant
Eq.~\ref{eq:Bjerrum_constant} reflects the discrete nature of both
individual interfacial charges and counterions and encodes their
mutual correlations. The main predictions of TC theories are
\cite{Travesset2006}:
\begin{itemize}
    \item Counterions bind with a fixed stoichiometric
    ratio to interfacial charges.
    \item The binding constant are electrostatic in origin and
    given by Eq.~\ref{eq:Bjerrum_constant} (in some situations,
    corrections need to be included \cite{Travesset2006}).
    \item The pair distribution functions satisfy the following
    relations for $r<<l_B |Q_{I}|q_{+}^{C}/2$, where C stands for counterion charge and I for
    interfacial charge:
    \begin{eqnarray}
    \label{eq:g_r_Bjerrum_O}
    g_{\rm CI}(r) &\approx& \exp(-q_{+}^{\rm C}Q_{I} \frac{l_B}{r}  )
    \ \\
    \label{eq:g_r_Bjerrum_Ba}
    g_{\rm CC}(r) &\approx& 0 \ ,
    \end{eqnarray}
$Q_I$ is assumed to be the charge of the interfacial molecule to
which the counterion is bound (for example, $Q_I=-1$ for PS$^-$
membranes). We stress that the condition $r<<l_B
|Q_{I}|q_{+}^{C}/2$ is an upper bound to the validity of the
correlation functions, see \cite{Travesset2006} for a more
detailed discussion.
\end{itemize}
In the language of Eq.~\ref{eq:inv_general}, the above description
corresponds to a free energy gain $\mu_{corr}=-k_BT\ln(K_I v)$,
where $v$ is a volume that depends on the units on which the
binding constant $K_I$ is defined.

\subsection{Electrostatic complexes or complexation correlation (CC) regime}

In this regime the Stern layer consists of electrostatic complexes
formed by several counterions, interfacial charges and water
molecules in equilibrium with the bulk solution. This regime will
be denoted as the complexation correlation (CC) regime herein and
will be illustrated with a concrete example in the next section.
It is characterized by:
\begin{itemize}
    \item Counterions generally bind to many interfacial charges.
    \item There are correlations among counterions and with interfacial charges, that is,
    the pair distribution function for both counterion-interfacial groups and bound counterion-counterion ions
    show peaks.
    \item The pair distribution function $g_{\rm{C I}}$, where
    I is an interfacial atom and C a counterion bound to it has a
    peak at $r_0$, the sum of the crystallographic radius for C
    and I atoms, with a width $l_0$ ($l_0/r_0<<1$) reflecting strong binding.
    Also, for $r<r_0+r_C$ (where $r_C$ is the crystallographic radius of the counterion
    C) it is $g_{\rm{C C}}(r)\approx 0$, reflecting that the complex
    staggers positive and negative charges.
\end{itemize}
Several approximations enable quantitative estimates within the CC
regime. In the general case, the complexes consist of a
distribution of patches with different sizes and compositions,
thus resulting in a distribution of correlation free energies for
counterions. We make the approximation that an average potential
of mean force $V_{MF}(r)$ for binding of a counterion to the
complex can be defined. From this potential of mean force, an
average binding constant is obtained by generalizing the binding
constant of the TC case, yielding the expression
\begin{equation}
\label{eq:Bjerrum_General}
    K_I^{CC}=c_{g}\int_{r_0}^{d_{Max}} dr 4\pi r^2 \exp(-\beta V_{MF}(r)) \
    ,
\end{equation}
where and $d_{Max}$ is a cut-off, which can be chosen as the
minimum of density $n(r)\sim r^2 \exp(-\beta V_{MF}(r))$,
following the same prescription as Bjerrum \cite{Robinson:1959}.
In general, a derivation of $V_{MF}$ requires a knowledge of the
chemical structure of the interfacial molecules, which makes its
evaluation a difficult task. We estimate the potential of mean
force from the observation that the binding to the complex is
electrostatic so we approximate the potential of mean force as
\begin{equation}
\label{eq:VMFeff} V_{MF}(r)\approx
\frac{q_{+}^CQ_I^{eff}}{\varepsilon r}
\end{equation}
where $r\geq r_0$ is the distance from the counterion to the
nearest neighbor interfacial groups and $q_{+}^C$ is the charge of
the counterion. The only parameter in this potential is
$Q_I^{eff}$, the effective charge of the binding site. Combining
Eq.(\ref{eq:VMFeff}) with Eq.(\ref{eq:Bjerrum_General}) provides
the generalization of Eq.(\ref{eq:Bjerrum_constant}) to the CC
regime \footnote{This expression is equivalent to
Eq.~(\ref{eq:Bjerrum_General}) only in the limit $\frac{|q_{+}
    Q_{-}^{eff}|l_B}{r_0}>> 1$, which is generally satisfied
    within the CC regime}
\begin{equation}
\label{eq:Bjerrum_constant_general}
    K_I^{CC}= c_{g} 4 \pi (|q_{+} Q_{-}^{eff}| l_B)^3 G(\frac{|q_{+}
    Q_{-}^{eff}|l_B}{r_0}).
\end{equation}
For this expression to be predictive, a prescription to compute
$Q_I^{eff}$ is required. This is obtained from the self-consistent
equation (which generalizes expressions
Eq.~(\ref{eq:g_r_Bjerrum_O}) and (\ref{eq:g_r_Bjerrum_Ba}))
\begin{eqnarray}
    \label{eq:g_r_Bjerrum_O_CC}
    g_{\rm CI}(r) &\approx& \exp(-\beta V_{MF}) \approx \exp(-q_{+}^{\rm C}Q_{I}^{eff} \frac{l_B}{r}  )
    \ \\
    \label{eq:g_r_Bjerrum_Ba_CC}
    g_{\rm CC}(r) &\approx& 0 \ ,
    \end{eqnarray}
expected to be valid for $r<d_{Max}$. The value of $Q_I^{eff}$ can
be obtained, for example, from MD simulations where the simulated
pair distribution function is fitted to the form
Eq.~\ref{eq:g_r_Bjerrum_O_CC} leaving $Q_{I}^{eff}$ as the only
fitting parameter. An example illustrating this case is discussed
next.

\section{The CC regime: The case of phosphatidic acid}
\label{SECT__Sim}

In this section we describe new MD simulation results (technical
details have been provided in \cite{Faraudo2006,Faraudo2006b}) of
an interface consisting of a phosphatidic acid lipid domain in
contact with a ionic solution containing BaCl$_2$. This system
provides a concrete realization of the CC regime and will be used
to discuss its most salient features.

\subsection{Description of simulations}

In our MD simulations, we have considered a monolayer with 100
DMPA$^{2-}$ phospholipids at close packing (molecular area
$\approx 41$ \AA$^2$) in contact with 50 divalent counterions
(Ba$^{2+}$) and added salt (100 BaCl$_2$). This system is
particularly suited for this study because experimental studies
report charge inversion \cite{Vaknin2003} and DMPA and other
lipids with the phosphatidic acid head group play a fundamental
role in a wide range of biological processes \cite{Wang2006}.

\begin{figure}
\includegraphics[width=8cm]{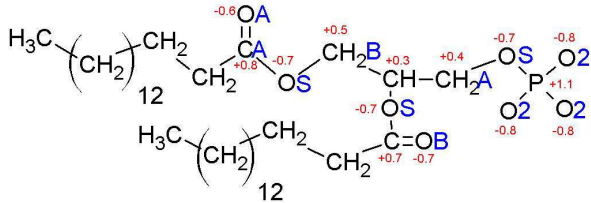}
\caption{Chemical structure of the DMPA molecule and its
assignation of electrical charges. OA,OS,O2,OB label different
oxygen atoms and CH$_2$A and CH$_2$A different hydrocarbon groups.
These distinctions are useful to characterize the different
binding sites for mobile ions for DMPA molecules.}
\label{fig:DMPA}
\end{figure}

The structure of the DMPA$^{2-}$
(1,2-dimyristoyl-\textit{sn}-glycero-3-phosphatidic acid)
phosphilipid molecule is given in Figure \ref{fig:DMPA} together
with its charge attributions following the AMBER force fields as
described in \cite{Smondyrev1999}. Water was included explicitly
within the SPC/E model and all simulations where carried with the
DLPOLY2 simulation package \cite{Forester2005}. Technical details
and a more extended analysis on other aspects of the simulations
can be found in \cite{Faraudo2006}. In this paper, we just provide
the results related to charge inversion and ion distributions.

\begin{figure}
\includegraphics[width=8cm]{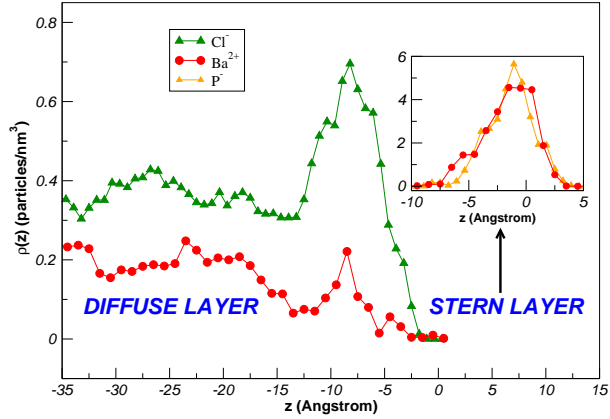}
\caption{(color online:) Number density distribution $\rho(z)$ of
Cl$^{-}$ (squares), Ba$^{2+}$ (circles), P (triangles) as a
function of the distance from the interface $z$. For the sake of
clarity, we show only a partial interval of the diffuse layer. The
Stern layer number density distributions are shown on the inset.
The position $z=0$ is defined so that it corresponds to the
maximum of the phosphate number density distribution.}
\label{fig:density}
\end{figure}

\begin{figure}
\includegraphics[width=8 cm]{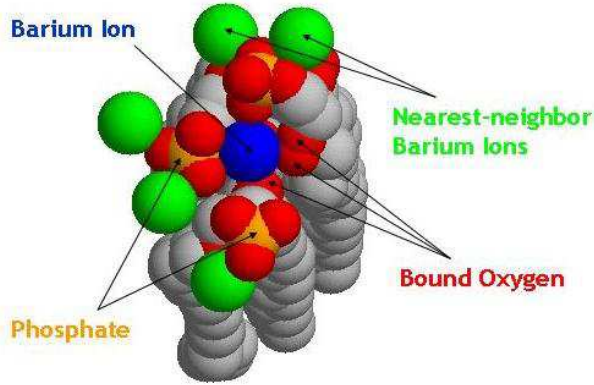}
\caption{(Color online:) Snapshot showing a Ba$^{2+}$ ion bound to
8 oxygens and 3 DMPA molecules.} \label{fig:Bind}
\end{figure}

\subsection{Complex formation and charge correlations within the Stern layer}

As shown in Fig.~\ref{fig:density}, the diffuse layer shows an
excess of negative charge (Cl$^{-}$) in the immediate vicinity of
the interface and a depletion of bulk counterions, which provides
conclusive evidence that the Stern layer has an overall positive
charge. Defining a counterion as bound if it has a DMPA oxygen
within its first coordination shell, we obtain for the number of
bound Ba$^{2+}$ $N^{\rm bound}_{\rm{Ba}}=1.065(10) N_{\rm{DMPA}}$.
Hence, the negative charge of the DMPA$^{2-}$ is overcompensated
by the counterions leaving a positive interfacial charge of
$q=0.13 e$ per phospholipid.

\begin{figure}
\includegraphics[width=8cm]{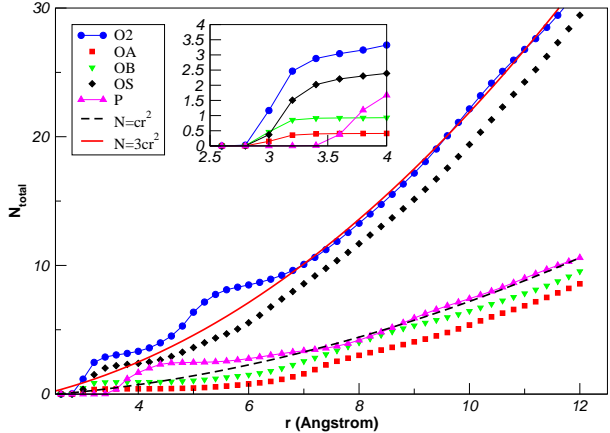}
\caption{(Color online:) Total number of atoms $N_{\rm total}$
within a distance $r$ from a given Ba$^{2+}$ ion. The coefficient
is $c=\frac{\pi}{40}$. The symbols are atom types, defined
according to Fig.~\ref{fig:DMPA}. The dashed and solid lines are
$N_{\rm total}$ for an equivalent smeared uniform distribution.
The inset shows an enlargement of the region comprising $2.5-4$
\AA.} \label{fig:N_r}
\end{figure}

The snapshot in Fig.~\ref{fig:Bind} shows binding of a given
counterion to several DMPA molecules, forming a complex involving
the interfacial oxygens, phosphates, additional barium counterions
and water molecules (not shown in the snapshot). These complexes
induce inhomogeneities within the surface charge distribution. In
Fig.~\ref{fig:N_r} the cumulative number of atoms of a particular
type $N_{\rm total}$ within a distance $r$ of a given bound
Ba$^{2+}$ ion is shown. If the interfacial charge could be
approximated as a uniform charge density $
    N_{\rm total}^a=\frac{\pi n_a}{40} r^2\equiv n_a c r^2$
where $n_a$ is the number of atoms of type $a$ per DMPA molecule
(from Fig.\ref{fig:DMPA}, $n_{\rm OS}=3$, $n_{\rm P}=1$, etc..).
The actual distribution for $N_{\rm total}^a(r)$ is shown in
Fig.~\ref{fig:N_r}. Large deviations from uniform distribution are
found for small distances $r< 6.5$\AA, which implies strong
correlations among Ba$^{2+}$ ions and interfacial groups. The
inset in Fig.~\ref{fig:N_r} shows an enlargement of the region
between $2.5-4$ \AA, where deviations from the uniform density are
the largest. At larger distances, small deviations are still
visible up to 10\AA.

\subsection{Pair distribution functions within the Stern layer}

\begin{figure}
\includegraphics[width=8cm]{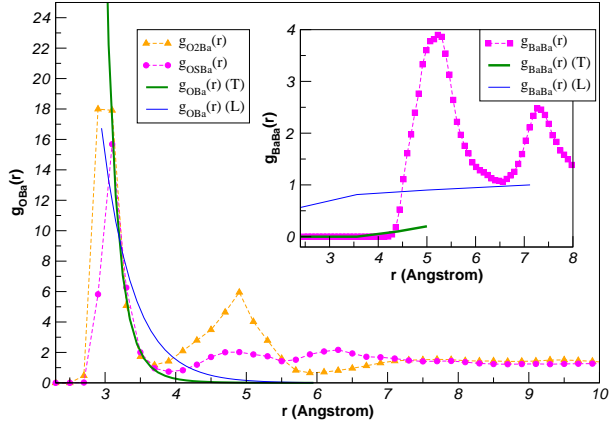}
\caption{(Color online:)Plot of the pair distribution functions
$g_{\rm O Ba^{2+}}(r)$ as a function of $r$, for O2 and OS oxygens
(see Fig.~\ref{fig:DMPA}). The inset shows the $g_{\rm Ba^{2+}
Ba^{2+}}(r)$ pair distribution. The result of CC predictions with
$Q_I^{eff}\approx -4$ Eq.~(\ref{eq:g_r_Bjerrum_O_CC}) is shown as
a solid line (CC) and LC predictions as thinner solid lines (LC).}
\label{fig:g_r}
\end{figure}

Results for the pair distribution functions $g_{\rm OBa^{2+}}$,
where O is one of the different oxygen atoms from DMPA$^{2-}$ (see
Fig.~\ref{fig:DMPA}) are shown in Fig.~\ref{fig:g_r}. The most
salient feature is a sharp peak located at a distance around $3$
\AA, which corresponds to the sum of the crystallographic radius
of oxygen and Barium (see Ref.~\cite{Israelachvili2000}),
confirming the strong binding among Ba$^{2+}$ and O. The second
peak and the barely visible third peak reflect correlations among
nearest neighbors phospholipids, which define a near crystalline
structure with lattice constant $6.8$\AA.

The simulations results for $g_{\rm OBa^{2+}}$ allow to obtain
$Q_I^{eff}$ as described previously with the fit shown in
Fig.~\ref{fig:g_r} (CC case), leading to an effective charge
$Q_I^{eff}\approx-4$. Bound counterions are tightly bound to 6-8
DMPA oxygens, so the charge in the immediate vicinity of a
Ba$^{2+}$ ion is between $-4.8e$ and $-6.4e$, which is
significantly lower than the obtained value $Q_I^{eff}\approx-4e$.
We regard this difference as an screening effect due to
neighboring positive charges. The value $Q_I^{eff}$ differs
substantially from the nominal charge of a single DMPA
phospholipid (which is $-2e$), thus implying that binding sites
with an effective local charge more negative than that of the
single molecule have been created at the interface.

Further understanding on the magnitude of the screening effects is
obtained from the analysis of the position of the first peak
(located at the sum of the crystallographic radius of Ba$^{2+}$
and O). A rough estimate for this peak position is obtained by
assuming that a Ba$^{2+}$ ion is bound to an O atom and ignoring
any other further away charges, leading to
$d_0=\left(\frac{48\sigma\varepsilon}{2|q_{-}|e^2}\right)^{1/11}\sigma\sim
2.7$\AA, where $\varepsilon$ and $\sigma$ are the values for the
Lennard-Jones contribution of the O-Ba interaction used in
simulations (see \cite{Faraudo2006}). This value is smaller by
10$\%$ from the one obtained in the simulations because of the
screening effects of nearby positive charges, which are ignored in
the estimate and repel the Ba$^{2+}$ from the O center. We include
those by assuming that there is a positive charge of valence
$q_{+}$ at a distance $d$ behind the Ba-O pair, which brings a
correction to the distance $d_0$ by $\Delta d_0=\frac{4 q_{+}
e^2}{\kappa(r_0+d)^2}$, where
$\kappa=11\left(\frac{2|q_{-}|e^2}{48\varepsilon\sigma}\right)^{3/11}\frac{2|q_{-}|e^2}{\sigma^3}$.
For the cases where the `repelling' charge is a P, $d=2$ \AA \ and
$q_{+}=1.3$, while if it is a pair of protons, $d=0.58$\AA \ and
$q_{+}=0.84$, both cases giving the $3$\AA \ quoted above. This
calculation shows that the leading contribution to the first peak
comes from the nearest neighbor O atoms, while the further away
positive charges can be included as a perturbation, very similarly
as in the discussion of the value of $Q_I^{eff}$.

The Ba-Ba pair distribution function is shown in the inset of
Fig.~\ref{fig:g_r}. It is quite apparent that Ba$^{2+}$ ions
strongly repel each other, with the first nearest neighbor
Ba$^{2+}$ being as far as 5 \AA \ away. The peaks of the  $g_{\rm
Ba Ba}$(r) distribution function can be qualitatively predicted
from the structure of the DMPA molecules. The average distance
among P groups in DMPA at molecular area $A_M=40$ \AA$^2$ is given
by $d_{PA}=\sqrt{\frac{2 A_M}{\sqrt{3}}}=6.8$ \AA. Therefore, we
expect a peak for  at distances $< d_{PA}$, which reflects binding
of two counterions to the same head group and other weaker (as
counterions become less correlated) peaks at distances $>d_{PA}$
reflecting Ba$^{2+}$ ions bound to a nearby DMPA. The position of
the first peak can be estimated more precisely from the
observation that two Ba$^{2+}$ ions bound to the same DMPA must
have O atoms, and have to be separated by the phosphate group,
thus providing a distance $d_0\approx d^{\prime}
\cos(109/2)\approx 5$ \AA, where $d^{\prime}$ is the Ba-P distance
on binding. Furthermore, this first peak implies a second peak in
the $g_{\rm Ba O2}$ function at about the same position, as the O2
oxygens are covalently attached to the P groups.

\subsection{PA domains provide an example of the CC regime}

From the previous discussion it is quite apparent that PA domains
provide an explicit example of charge inversion by CC. We now
analyze whether such results can be described as either LC or TC
regimes.

The results from the MD simulations show the critical role played
by the interfacial oxygens in providing binding sites for
counterions, as revealed by the snapshot shown in
Fig.~\ref{fig:Bind} and more quantitatively by Fig.~\ref{fig:N_r},
which shows the inaccuracies of approximating the surface charge
by a continuum. Therefore, the basic hypothesis for LC theories do
not hold in this case.

The pair distribution of the OCP describes $g_{\rm Ba^{2+}
Ba^{2+}}$ within LC and is tabulated in Ref.~\cite{Totsuji1978}
and shown in Fig.~\ref{fig:g_r}. OCP predicts that the probability
of finding Ba$^{2+}$ ions in proximity is much higher than it is
found in our simulations. Furthermore, counterions at the
interface are not in a liquid state, but are bound to the
interface with occasional exchange with the bulk. The ion
distribution within LC is described by Eq.~(\ref{eq:g_r_LC_O}),
which is in disagreement with MD results as shown in
Fig.~\ref{fig:g_r}. In conclusions, the results of the MD
simulations are not consistent with charge inversion as described
within LC theories.

The snapshot in Fig.~\ref{fig:Bind} shows that a typical
counterion is bound to several DMPA molecules, and that
counterions may share the same DMPA as binding sites. Therefore,
there is no clear evidence of binding with a fixed stoichiometric
ratio, which is one of the key assumptions within TC theories. The
pair distribution between Ba$^{2+}$ ions should satisfy
Eq.~\ref{eq:g_r_Bjerrum_Ba}, and as shown in Fig.~\ref{fig:g_r}
this equation is satisfied within the first 4.5 \AA, but a strong
peak, indicating correlations among nearby bound Ba$^{2+}$ ions is
observed at a distance of 5\ \AA. Furthermore, according to TC
theories, the first peak of the pair distribution between O and
Ba$^{2+}$ should be described by Eq.~\ref{eq:g_r_Bjerrum_O} with
the nominal charge of DMPA ($Q_I=-2$) while as described in the
simulations, a value $Q_I^{eff}\approx-4$ is obtained. Although
the predictions within TC models are able to reproduce some of the
features observed, particularly at short distances $<4.5$\AA,
several predictions are in clear disagreement and TC theories do
not provide a satisfactory description of the MD simulations.

\subsection{Properties of PA domains as a CC regime}

From the previous discussion we can now compute the potential of
mean force Eq.(\ref{eq:VMFeff}) and the average binding constant
from Eq.~\ref{eq:Bjerrum_General}. Assuming values for the steric
coefficient $c_g$ between 1 (no steric effects) and 1/10 (strong
steric reduction) we obtain a binding constant in the range
$K_I^{eff}\approx 10^6-10^{7}$ M$^{-1}$. This very large value
shows a strong affinity of DMPA$^{2-}$ for divalent cations.

Implicit in all our simulations is that DMPA$^{2-}$ is doubly
charged. We discuss this point in more detail, as DMPA can exist
as DMPA$^{-1}$ or even as a neutral entity. We assume that DMPA
has the pK$_a$ values of phosphoric acid, a first pK$_{a}^1=2.1$
and a second pK$_{a}^2=7.1$ \cite{Atkins}. The validity of using
bulk pK$_{a}$ values to interfaces has been shown in
\cite{Travesset2006,Bu2005a,Bu2005b}. The critical Ba$^{2+}$
concentration, c$_{crit}$ where DMPA becomes doubly deprotonated
is obtained when the free energy for Ba$^{2+}$ binding is the same
as the free energy of a proton being transferred to DMPA$^{2-}$.
This leads to $\ln(K_B$c$_{crit})=\ln(10^{pH-pK_a^2})$ or
c$_{crit}\sim 10^{-7}/c_g$ M. Even if the geometric correction
$c_g$ increases c$_{crit}$ by an order of magnitude, the presence
of divalent ions should doubly-deprotonate DMPA already in
extremely dilute regimes, and charge inversion should immediately
follow ( $c_{inv}\sim 1/K_I$). For DMPA domains, deprotonation and
charge inversion occur simultaneously. These results illustrate
how correlations affect many other processes such as proton
release and transfer.

\subsection{PA domains with monovalent ions}

We now briefly discuss how results are modified in the presence of
monovalent ions. In this case, the binding constants
Eq.~\ref{eq:g_r_Bjerrum_Ba} give $c_{inv}> 1M$ (where the dilute
regime is no longer valid). Generally, the resulting binding
constant predict a small binding fraction of monovalent ions, even
more so for larger ions such as Cs$^{+}$. For the case of DMPA,
the monovalent ions fully release the first proton (becoming
DMPA$^{-}$) only at concentrations larger than $\sim 10$
\cite{Bu2005a,Bu2005b} mM but contrary to the situation with
divalent ions, the balance of free energies does not strip off the
second proton. In general, we expect that solutions of monovalent
ions are systems without LC regimes and weak or negligible TC or
CC regimes, thus following mean field theory very accurately in
dilute regimes, as shown recently from X-ray experiments
\cite{Bu2005b,Bu2005a,Luo2006} and emphasized earlier by
McLaughlin \cite{McLaughlin1989}.

\section{Comparison of the different regimes with experimental results}
\label{SECT__Discussion}

In previous sections we identified three correlation regimes and
provided a detailed description of the CC regime by analyzing a
particular case. We now provide specific criteria to identify the
expected range of validity for each regime and discuss a
representative experimental example.

\begin{figure}
\includegraphics[width=8cm]{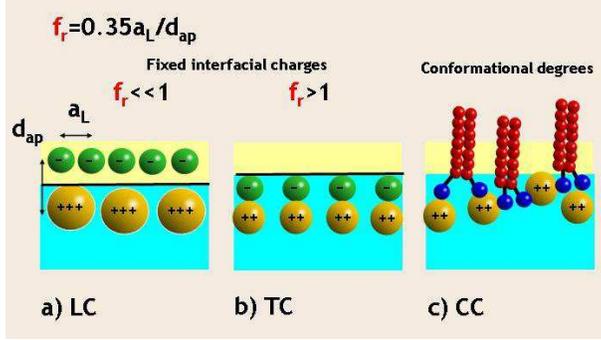}
\caption{Summary of the range of validity of the different
theories. With interfacial molecules without conformational
degrees of freedom, the transition from LC to TC depends on the
parameter $f_r$. When the interfacial molecules have
conformational degrees of freedom, complexation (CC regime) may
occur.} \label{fig:Scenarios}
\end{figure}

The appearance of the different regimes depends on whether the
charges at the interface are fixed or have conformational degrees
of freedom and whether the bound counterions ``see'' the discrete
interfacial charges or a smeared interface, which is controlled by
the parameter
\begin{equation}\label{eq:f_r_param}
    f_r=0.35 a_L/d_{ap} \ .
\end{equation}
The prefactor 0.35 corresponds to a hypothetical situation where
both interfacial charges and counterions are forming crystalline
structures \cite{Travesset2006}, but we use this factor in any
other situation (a factor of 1 severally overestimates the role of
discrete charges, while a factor of $1/(2\pi)$, proposed in, for
example \cite{Israelachvili2000}, underestimates discreteness
effects). In the above formula $a_L$ is the distance among
interfacial charges and $d_{ap}$ is the distance between a
counterion and a charged interfacial group (see
Fig.~\ref{fig:Scenarios}). For $f_r<<1$, the counterions ``see''
the smeared interface, while for $f_r>1$ they ``see'' the discrete
interfacial charges (We emphasize that as shown in
\cite{Travesset2006}, discrete charge effects are already present
for $f_r$ not much larger than 1, so we keep the condition as
$f_r>1$, instead of $f_r>>1$). We now discuss each regime in turn.

\subsection{LC theories}

For LC theories to apply it is basically necessary that
interfacial charges may be approximated as a continuum. This can
be achieved if 1) the interfacial charges are fixed (do not have
conformational degrees of freedom) and 2) the condition $f_r<<1$
(see Eq.~(\ref{eq:f_r_param})) is satisfied.

A strong experimental candidate for the LC regime is provided by
solid surfaces in contact with different trivalent and tetravalent
ions with large radius ($r_{+}\sim 4-4.5$\AA \ )
\cite{Besteman2004,Besteman2005}. In those experiments, condition
1) was trivially satisfied by the nature of the interface, while
from the values reported for the bare surface charge, it is found
$f_r\sim 0.08-0.4 <<1$, so condition 2) is also met. Indeed, the
detailed experimental results show general agreement with LC
theories of Shklovskii and collaborators
\cite{Shklovskii:99,Grosberg2000}. For example, the concentration
at which charge inversion appears ($c_{inv}$) is very well
predicted from Eq.(\ref{eq:inv_general}) and the correlation
energy described in Sect.~\ref{SECT__Theory}A. Recent experiments
\cite{Lemay2006} show that charge inversion disappears at higher
trivalent ion concentrations, in marked disagreement with the
predictions in \cite{Grosberg2000}, but in agreement with the LC
theory of Pianegonda et al. \cite{Pianegonda2005}, who include the
effect of bulk Bjerrum pairing in the chemical potential of the
bulk solution.

It is possible that the results in \cite{Besteman2005} can be
described by a TC theory? The general trends of $c_{inv}$
regarding its dependence on dielectric constant and multivalent
concentration are in agreement with the binding constant
Eq.~\ref{eq:Bjerrum_constant}. The strongest evidence against
charge inversion as described within the TC regime comes from the
observed dependence of $c_{inv}$ with increasing bare surface
charge, which according to the TC regime is given by $c_{inv}\sim
1/K_I$, that is, independent of surface charge. Understanding the
observed dependence of $c_{inv}$ on bare surface charge within the
TC regime could only be explained by appealing to complex binding
scenarios where  counterions bind to several interfacial charges,
which in view of the good agreement within LC theories, does not
seem justified. We point out, however, that similar experiments
for divalent (Ca$^{2+}$ and Mg$^{2+}$) ions show that $c_{inv}\sim
0.3$M, in disagreement with LC theories, which predict
$c_{inv}\sim 0.01$M. We point out that Ca$^{2+}$ and Mg$^{2+}$
have a small radius leading to $f_r\sim 2$, and indeed, in this
case the TC regime predicts $c_{inv}\sim 1/K_I^{TC} \approx 0.1$M.
It is quite possible that the experiments reported in
\cite{Lemay2006} for divalent ions provide an example for the TC
regime. Further experiments or simulations will be required to
establish this point.

\subsection{TC theories}

TC theories require that 1) that interfacial charges do not have
conformational degrees of freedom and 2) $f_r=0.35a_L/d_{app} >
1$.

A clear experimental candidate for charge inversion for the TC
regime is provided by PS membranes in a solution containing
Ca$^{2+}$ ions \cite{McLaughlin:1981a}. It is found that $f_r\sim
4$ so condition 2) is satisfied. It is not entirely obvious that
condition 1) is satisfied, as PS lipids have significant
conformational degrees of freedom. The experiments in
\cite{McLaughlin:1981a} where done with brain PS, which contain
different degrees of unsaturated hydrocarbon chains that might
leave PS head groups sufficiently separated to prevent
complexation from taking place. More detailed analysis, which can
be provided from accurate MD simulations will be required to
assess the degree of complexation found in these systems. The
binding constant computed from Eq.~\ref{eq:Bjerrum_constant} for
Ca$^{2+}$ give $K_I^{TC}\approx 10$M$^{-1}$, which allows to
reproduce the values of contact value potential extracted from
electrophoretic measurements over five decades in salt
concentration \cite{McLaughlin:1981a}without any fitting
parameters. Furthermore, results for Na$^{+}$ are also reproduced
from Eq.~\ref{eq:Bjerrum_constant} without fitting parameters. It
is therefore quite apparent that the experimental results of
\cite{McLaughlin:1981a} are described by TC theories
\cite{Travesset2006} with surprising accuracy. It is remarkable
that the same binding constants can be used to predict ionic
distributions at higher surface density, such as in Langmuir
monolayers \cite{Bloch1990,Travesset2006} (with some corrections,
as discussed in \cite{Travesset2006}) and in diluted PS membranes
\cite{Huster2000} with considerably lower surface charge density.
These observations are in full agreement with the expectations
from the TC regime, since Eq.(\ref{eq:Bjerrum_constant}) predicts
that the binding constant does not depend on the surface charge
density $\sigma_0$.

Is it possible that the PS membranes are described by LC theories?
LC theories would predict a much lower $c_{inv}$ (in the mM
regime) and $\zeta$ potential values should exhibit some
corrections due to the presence of the two dimensional correlated
liquid. Furthermore, the fact that binding constants are
independent of surface charge provides strong evidence against PS
membranes being described by LC theories. In general, amphiphilic
interfaces show a strong dependence on the nature of the
interfacial molecule, which makes it generally incompatible with
the LC regime. For example, DMPA monolayers with a surface charge
$\sigma_0\simeq -e/20 $\AA $^2$ show charge inversion in presence
of $\mu$M concentrations of BaCl$_2$ \cite{Vaknin2003} but fatty
acids, with basically the same surface charge, do not exhibit
charge inversion \cite{Kjaer:1989}.

Regarding the description of PS membranes within CC theories, in
\cite{McLaughlin:1981a} it was assumed that divalent ions bind in
a 1:1 stoichiometric ratio, while in \cite{Huster2000}, a 2:1
ratio was assumed, both experiments giving good results.
Furthermore, the results in \cite{McLaughlin:1981a} show some
specificity in ionic type, which is slightly larger than predicted
from TC theories. Although TC describe the experiments with
surprising accuracy, it is therefore likely that more detailed
studies will reveal some degree of complexation.

\subsection{CC theories}

Complexation requires that interfacial groups have conformational
degrees of freedom. X-ray reflectivity experiments with monolayers
of DMPA at molecular area $A_M=40$\AA$^2$ report charge inversion
\cite{Vaknin2003} at the mM regime for Ba$^{2+}$, and some
indications that it may extend at least up to the $\mu$M range
\cite{Vaknin2006c}. Recent results provide charge inversion within
the $\mu$M range for La$^{3+}$ \cite{Pitler2006}. If these results
were described by TC, $c_{inv}$ would be in the $10^{-3}$M range
for divalent ions and in the $5\cdot 10^{-4}$M for the trivalent
ions. Clearly, the observed charge inversion takes place at much
lower concentrations and is consistent with the predictions of CC
discussed in the previous section.

A more stringent test to assess the prediction of the CC theory
described in the previous section would require 1) to assess the
number of interfacial O surrounding a given Ba$^{2+}$ and 2)
establishing the presence of a diffuse layer of co-ions (Cl$^{-}$,
etc..) next to the surface. Both predictions can be realistically
validated from X-ray studies. X-ray spectroscopy from recently
developed surface sensitive EXAFS \cite{Vaknin2006b} has provided
a detailed description of the oxygen atoms bound to Cs$^{+}$ ions
next to charged interface, so the same technique could be applied
to Ba$^{2+}$ solutions in order to elucidate point 1). Regarding
point 2), if the amount of charge inversion is substantial, use of
heavy ions such as I$^{-}$ (as opposed to Cl$^{-}$), which have
resonances in the X-ray region, could be detected by anomalous
X-ray surface sensitive scattering \cite{Vaknin2003} combined with
fluorescence methods.

\section{Conclusions}\label{SECT__Conclusions}

Charge inversion always results from a favorable correlation free
energy (Eq.~\ref{eq:inv_general}). We identified three different
regimes leading to charge inversion, schematically illustrated in
Fig.~\ref{eq:inv_general}) and provided specific criteria to
theoretically identify each different regime (also shown in
Fig.~\ref{eq:inv_general}), further illustrating it with a
representative experimental example.

We have provided a detailed discussion of the CC regime using
recent MD simulations of phosphatidic acid domains
\cite{Faraudo2006} and showed how processes such as proton
transfer and release are intimately related to correlations.
Furthermore, our results show the critical role played by strongly
electronegative atoms (the oxygens), which create binding sites,
as shown in Fig.~\ref{fig:Bind}, leading to the formation of
highly cohesive complexes. This observation emphasizes the
necessity of including atomic details in investigations of charged
monolayers and membranes. Future ab-initio calculations, with more
rigorous inclusion of polarizability effects, may provide more
precise insights into these issues. Our results show possible
pitfalls of excessive coarse-grained theories and simulations.
Considering a model where DMPA consists of single divalent point
charge at the head group would be unable to generate the kind of
complexes shown in Fig.~\ref{fig:Bind}. In this hypothetical
coarse-grained model, divalent counterions would most likely bind
by TC in a stoichiometric ratio of 1:1 and charge inversion could
not follow.

Although our MD simulations have only discussed the case of DMPA,
our study can be used to predict ion distribution and charge
inversion in many other systems. Fatty acids, for example, with a
surface charge very close to DMPA do not exhibit charge inversion
\cite{Kjaer:1989}. Fatty acids only have two oxygens per molecule
available for binding and binding constants become on the order $1
$M$^{-1}$ \cite{Travesset2006}. A similar situation is observed
for DHDP (Dihexadecyl hydrogen-phosphate)\cite{Bu2005b}, which
does not exhibit charge inversion. On the other hand, we predict
that charge inversion should be a common effect in membranes of
biologically relevant charged phospholipids such as phosphatidyl
serine (PS), phosphatidylinositol or glycolipids such as the
gangliosides, which contain more than 10 oxygens per molecule and
can become multiply charged. Detailed quantitative predictions for
these lipids will be presented elsewhere.

The role of hydration sheaths on binding has not been discussed in
this paper. Although a detailed analysis is reported elsewhere
\cite{Faraudo2006}, our simulation results show that Ba$^{2+}$
ions lose roughly half of their hydration sheaths upon binding, so
that the distance between bound ions and oxygens is given by the
sum of their crystallographic radius. It is expected that this is
a general result, as with very few exceptions, all negatively
charged interfaces consist of oxygen atoms, and as exemplified
from Fig.~\ref{fig:DMPA}, oxygen charges are basically the same
whether the oxygen is from water or within an interfacial group,
and therefore trading a water for an interfacial oxygen is
entropically favored, similarly as in mechanisms for ion
selectivity in ion channels \cite{Alberts}. Our results also
extend to other ions such as Ca$^{2+}$ or Sr$^{2+}$, which share
the same electronic structure with Ba$^{2+}$ but with a
considerably smaller crystallographic radius, which should enhance
charge inversion by allowing the ions to get closer to the O atoms
and increase their binding. Other divalent ions like Cd$^{2+}$,
which do not have the electronic structure of a noble gas, bind
more strongly than Ba$^{2+}$ or Ca$^{2+}$, but this binding is
mainly covalent \cite{Travesset2006}. This situation is usually
referred as specific binding \cite{Besteman2005} and can be
described by a phenomenological binding constant encoding the free
energy of the covalent bond. In this situation, $K_L$ cannot be
estimated theoretically (with a formula such as
Eq.~\ref{eq:Bjerrum_constant}) and it is independent of
environmental variables such as the dielectric constant of the
solution, the charge surface density of the interface, etc..

In conclusion, our results show that charge inversion is a common
effect in charged interfaces in contact with a solution of
multivalent ions, but it may have many different origins. How our
results may extend to other charged systems such as proteins or
biopolymers like DNA or actin remains the subject for future work.

{\bf Acknowledgements}

We acknowledge D. Vaknin for his many insightful remarks and C.
Lorenz and M. Losche for discussions. JF acknowledges many
discussions with R. Kjellander at the XIIIth International
Conference on Surface Forces in Moscow. A.T. acknowledges many
discussions with S. Lemay, T. Nguyen, F. Pincus and B. Shklovskii
at the Aspen center for physics. This work is supported by NSF
grant DMR-0426597, the Spanish Government Grant No.
FIS2006-12296-C02-01, the UAB grant IEME2005-46 and partially
supported by DOE through the Ames lab under contract no.
W-7405-Eng-82. The authors thankfully acknowledge the computer
resources, technical expertise and assistance provided by the
Barcelona Supercomputing Center - Centro Nacional de
Supercomputaci\'on.


\end{document}